\documentclass[acmtog, nonacm, balance = false]{acmart}
\AtBeginDocument{%
  }

\setcopyright{none}

\newtheorem{theorem}{Theorem}

\begin{document}

\title{A Sharp Conservative Offset for Marching Cubes}

\author{Alec Jacobson}
\email{jacobson@cs.toronto.edu}
\orcid{0000-0001-7438-5132}
\affiliation{%
  \institution{University of Toronto}
  \country{Canada}
}

\begin{abstract}
Given a grid-sampled 1-Lipschitz function $f:\mathbb{R}^3 \rightarrow \mathbb{R}$ (such as a signed distance function) sampled on a regular grid, 
we determine the 
sharp infimum $\sigma_\star$ such that for every
$\sigma > \sigma_\star$ the true zero-level set $f^{-1}(0)$ is contained in the marching cubes mesh for $f = \sigma$. For grids with spacing $h$, the sharp infimum is $\sigma_\star = \frac{\sqrt{3}}{2} h$. This result is proven \emph{despite} the planar marching cube faces potentially lying on either side of the level set of the trilinearly interpolated approximation of $f$ (which shares the same infimum). 
The proof instead constructs a convex lift of the corner samples and
shows that the high-side portion of each marching-cubes cell lies in
the projection of its $\sigma$-superlevel portion, which cannot contain
a point of $f^{-1}(0)$.
\end{abstract}

\maketitle

\section{Setup}
We consider a $1$-Lipschitz function $f : \mathbb{R}^3 \rightarrow \mathbb{R}$ where $f<0$ defines the inside solid and the region $f>0$ is called outside.
Let us sample $f$ onto nodes of a Cartesian grid $h\mathbb{Z}^3$ where $h > 0$ is the grid spacing. We write $f_v = f(v)$ for value sampled at vertex $v$ of the grid.
Let $\mathcal{M}_\sigma$ be the marching cubes mesh extracted at the $\sigma$ level set. In particular, for every edge of the grid $(a,b)$ marching cubes emits a vertex if the classification $f_a < \sigma $ does not match $f_b < \sigma$, placing it according to linear interpolation along the edge. The grid vertices at either end of such an edge are referred to as being on the low and high side, respectively. Then each cell connects its emitted edge-crossing vertices into triangles according to its case resolution \cite{LorensenC87}. There are many variants of marching cubes. 
We consider an edge-only marching-cubes variant: every output vertex
lies on a sign-changing grid edge, every triangle lies in its parent
cell, and the resolved mesh is a watertight, sign-respecting separator 
of the low and high grid vertices. In particular, the $\text{inside}(\mathcal{M}_\sigma)$ is the union of all per-cell low-side regions.

\begin{theorem}[Sharp enclosure offset]
\label{the:sharp}
If
\[
  \sigma> \sigma_\star = \frac{\sqrt{3}}{2}h,
\]
then
\[
  f^{-1}(0)\subset \text{inside}(\mathcal{M}_\sigma).
\]
The constant $\sigma_\star = \sqrt{3}h/2$ is sharp as an infimum over signed distance
functions to smooth solids.
\end{theorem}

\section{Irrelevance of trilinear interpolation's sharp offset}
We point out first that this is distinctly different from a similar statement.
Consider the function $t : \mathbb{R}^3 \rightarrow \mathbb{R}$ which is constructed as the trilinear interpolation of $f$'s vertex-values on the grid.
If we choose $\sigma>\frac{\sqrt{3}}{2}h$ then it is also provable that 
the $0$-level set of $f$ lies on the low-side of 
the $\sigma$-level set of $t$:
\[
  f^{-1}(0)\subseteq\{x:t(x)<\sigma\}.
\]

\begin{figure}
\includegraphics[width=\linewidth]{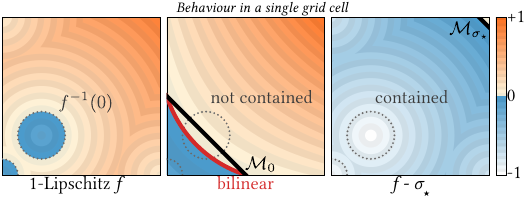}
\caption{\label{fig:2d} Left: a non-trivial $1$-Lipschitz function in a single cell. 
Middle: bilinear interpolation agrees with the 0-level set marching squares mesh at edge-crossings but not in the interior.
Right: regardless, the uniformly safe offset $\sigma_\star$ keeps the true 0-level set contained.
}
\end{figure}

Unfortunately, it is not clear that this is useful for proving Theorem~\ref{the:sharp}. Even when the vertices of the marching cubes mesh $\mathcal{M}_\sigma$ are chosen according to (tri-)linear interpolation, the emitted faces are planar and can (and almost always do) cut on either side of the piecewise-smooth trilinearly interpolated level-set $t^{-1}(\sigma)$ (see Fig.~\ref{fig:2d}).
The worst-case distance between the trilinearly interpolated isosurface and the marching cubes mesh is at least $h 2/\sqrt{3}$ (consider $f = xyz - \varepsilon$.
Bounding by the worst case distance would provide a conservatively larger $\sigma$ infimum for marching cubes, but 
we will now show that we can make it just as tight for marching cubes directly.

\section{Proof}
Without loss of generality, consider a grid cell $[0,h]^3$ and define $R \subseteq [0,h]^3$ to be the high-side portion of the cube bounded by cube-face pieces and triangles of the marching cubes mesh $\mathcal{M}_\sigma$ emitted in $[0,h]^3$. The solid $R$ is bound by planar faces connecting high-side cell vertices and vertices emitted by marching cubes at edge-crossings. Gather these vertices into a set $B$.
For a high-side cell vertex $v$ we have $f(v) \ge \sigma$. For a vertex $q$ emitted at crossing an edge  $(a,b)$ we have:
\[
q = (1-t) a + t b \quad \sigma = (1-t) f(a) + t f(b).
\]

Now let's define the convex set $P$ as
\[
x \in P \iff \exists \, \alpha \text{ such that } \begin{cases}
x = \sum_{i=1}^8 \alpha_i v_i, \\
\alpha_i \ge 0, \sum_{i=1}^8 \alpha_i = 1, \\
\sum_{i=1}^8 \alpha _i f(v_i) \ge \sigma,
\end{cases}
\]
where $v_i$ runs over all eight corners of the cell.
Each high-side cell corner $v_j$ lies in $P$ (with $\alpha_i = \delta_{ij}$).
Similarly, each edge-crossing vertex $q$ lies in $P$ (with $\alpha_a = (1-t)$ and $\alpha_b = t$).
Define for each $b \in B$ its corresponding $\alpha^b$.

As the boundary of $R$ is made up of planar faces bounded by edges connect the vertices in $B$, we know that $R \subseteq \text{conv}(B)$. Consequently, since $\text{conv}(B) \subseteq P$, we also have $R \subseteq P$. 

Now, assume there is a point $p \in [0,h]^3$ satisfying $f(p) = 0$, and (for contradiction) 
$p$ lies in the high side $R$ of the marching cubes mesh.

We can show that an appropriate $\alpha$ for $p$ exists.
Since $p \in \text{conv}(B)$ there are $\lambda_b \ge 0$ summing to $1$, such that: $p = \sum_{b \in B} \lambda_b b$. Then explicitly take $\alpha_i = \sum_{b \in B} \lambda_b \alpha^b_i$.
This establishes:
\[
p = \sum_{i=1}^8 \alpha_i v_i \quad \text{ and } \quad \sum_{i=1}^8 \alpha_i f(v_i) \ge \sigma.
\]

Since $f$ is $1$-Lipschitz and $f(p) =0 $ (or more generally $f(p)\leq 0$), 
\[
f(v_i) \leq f(p) + \|v_i - p\| \leq  \|v_i - p\|.
\]
Multiplying by $\alpha_i \ge 0$ and summing gives,
\[
\sum_{i=1}^8 \alpha_i f(v_i) \leq \sum_{i=1}^8 \alpha_i \|v_i - p\|.
\]
Then Jensen's inequality for the concave square-root function gives:
\[
\sum_{i=1}^8 \alpha_i \| v_i - p \| \leq \sqrt{
\sum_{i=1}^8 \alpha_i \|v_i - p\|^2
}.
\]
Let's treat $X$ as a random cube corner with probabilities $\alpha_i$. For a coordinate  $k$ 
\[
\mathbb{E}[X_k] = p_k, \quad X_k \in \{0,h\},
\]
so 
\[
\mathbb{E}[(X_k-p_k)^2] = p_k(h - p_k) \leq \frac{h^2}{4}.
\]
Therefore,
\[
\sum_{i=1}^8 \alpha_i \|v_i - p\|^2 = \sum_{k=1}^3 p_k (h - p_k) \leq \frac{3 h^2}{4}.
\]
Connecting it all together, a point $p$ in the high-side, can only exist if:
\[
\sigma 
\leq \sum_{i=1}^8 \alpha_i f(v_i) 
\leq \sum_{i=1}^8 \alpha_i \|v_i - p\|
\leq \sqrt{\sum_{i=1}^8 \alpha_i \|v_i - p\|^2}
\leq \frac{\sqrt{3}}{2} h.
\]

To show sharpness as an infimum, consider the signed distance to a small sphere placed  at the centre
of the cell $c$
with positive radius $r$:
\[
f(x) = \|x - c\| - r.
\]
The grid vertices have value $h \sqrt{3}/2 - r$. Therefore, for every $\sigma < h\sqrt{3}/2$, choose 
$0 < r  < h \sqrt{3}/2 - \sigma$. All vertices are on the high side and marching cubes emits no triangles, despite the true sphere $f=0$.

\subsection{Generalizations}
There are a variety of straightforward generalizations.
For anisotropic grid spacing, we can show $\sigma_\star = \sqrt{h_x^2 + h_y^2 + h_z^2}/2$.
For marching ``cubes'' in dimension $d$, we have $\sigma_\star = h \sqrt{d}/2$ (always half the cell diagonal).
If $f$ is $K$-Lipschitz then $\sigma_\star = h K \sqrt{3}/2$.
If we use root-finding at edge-crossing vertices $q$ so that $f(q) = \sigma$, then 
$\sigma_\star = h \sqrt{3}/2$ remains conservative and sharp (replacing the eight corners with $B$ directly).
%
Some variants of marching cubes (e.g., \cite{Chernyaev1995,Grosso16}) may insert an additional vertex in the interior of the cell: our proof extends so long as this vertex 
lies in $P$ or (sufficiently) that it is a convex combination of the edge-crossing vertices.

\section{Related work and applications}
The $1$-Lipschitzness of implicit functions has long been exploited in graphics and geometry processing.
In rendering, the value of $f$ is used as a safe radius for ray-casting 
and cell pruning \cite{KalraB89,Hart96,barbier25lipschitz}. These settings are in a way the inverse setup involving the same value. To 
reject an entire grid cell as not containing $f^{-1}(0)$, we can
evaluate $f$ at the center $c$ and verify that $|f(c)| > h \sqrt{3}/2$.
Naively flipping this around for our setting, the query point becomes \emph{any} probe point $p$ in the cell, so to ensure that $f(p)>0$, we 
would require at each grid sample vertex $f(v) > h \sqrt{3}$: loose by a factor of $1/2$. 

\begin{figure}
\includegraphics[width=\linewidth]{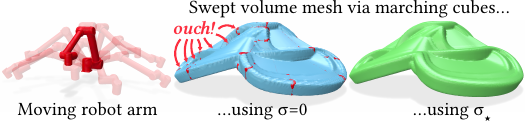}
\caption{
\label{fig:swept-volume}
Swept volume methods often work with implicit functions but rely on marching cubes for final extraction, but if that mesh intersects the moving object then safety guarantees evaporate.
}
\end{figure}

A sharp conservative offset is useful when marching cubes meshing occurs as a subroutine for tasks that require strictly maintaining a containment invariance. 
The unsigned distance contouring of \citet{Hou_2023} begins with a marching cubes mesh of a small offset, but makes no guarantee that the original 0-level set won't poke through if the offset \emph{mesh} is too small.
In the cage construction of \citet{sellan2021}, high-resolution meshed levelsets are the initial input to a decimation engine. In swept volume extraction, implicit functions are particularly useful to aggregate signed distances of a shape over its temporal motion \cite{SchroederLL94,SellanSwept2021}. But if the final extracted mesh is meant for safe downstream collision tasks, then a sharp and conservative offset is appreciated (see Fig.~\ref{fig:swept-volume}).

\bibliographystyle{ACM-Reference-Format}
\bibliography{references}

\end{document}